\documentclass{llncs}
\usepackage{amssymb}
\usepackage{amsmath}
\usepackage{times}
\usepackage{comment}
\usepackage{graphicx}

\newtheorem{observation}{Observation}
\title{On the read-once property of branching programs and CNFs of bounded treewidth}
\author{Igor Razgon}
\institute{Department of Computer Science and Information Systems, Birkbeck, University of London\\
           igor@dcs.bbk.ac.uk}
\begin{document}
\maketitle
\begin{abstract}
In this paper we prove a space lower bound of $n^{\Omega(k)}$ for 
non-deterministic (syntactic) read-once branching programs 
({\sc nrobp}s) on functions expressible as {\sc cnf}s with treewidth at most $k$ of their 
primal graphs. This lower bound rules out the possibility of fixed-parameter space complexity 
of {\sc nrobp}s parameterized by $k$.

We use lower bound for {\sc nrobp}s to obtain a quasi-polynomial separation between Free Binary Decision Diagrams
and Decision Decomposable Negation Normal Forms, essentially matching the existing upper bound introduced by Beame et al.
and thus proving the tightness of the latter.
\end{abstract}
\section{Introduction}
\subsection{Statement of results and motivation}
Read-once Branching Programs ({\sc robp}s) are a well known representation of Boolean functions. 
Oblivious {\sc robp}s, better known as  Ordered Binary Decision Diagrams ({\sc obdd}s),
are a subclass of {\sc robp}s, very well known because of its applications in the area of verification \cite{OBDD}.
An important procedure in these applications is transformation of a {\sc cnf} formula into an equivalent {\sc obdd}.
The resulting {\sc obdd} can be exponentially larger than the initial {\sc cnf} formula, however a space efficient
transformation is possible for special classes of functions. For example, it has been shown in \cite{VardiTWD}
that a {\sc cnf} formula with treewidth $k$ of its primal graph can be transformed into an {\sc obdd} of size $O(n^k)$.
A natural question is if the upper bound can be made fixed-parameter i.e. of the form $f(k)n^c$ for some constant $c$.
In \cite{RazgonKR} we showed that it is impossible by demonstrating that for each sufficiently large $k$
there is an infinite class of {\sc cnf} formulas with  treewidth at most $k$ whose smallest {\sc obdd} is of size at least 
$n^{k/5}$.


In this paper we report a follow up result (Theorem \ref{maintheor}) showing that essentially the same lower bound holds for
non-deterministic read-once branching programs ({\sc nrobp}s). \footnote{Throughout this paper, we assume the
read-once property to be \emph{syntactic}, that is applied to \emph{all} root-leaf paths of the considered branching programs.
See Section 2 for the exact definitions.}
In particular we show that there is a constant $0<c<1$
such that for each sufficiently large $k$ 
there is an infinite class of {\sc cnf} formulas of treewidth at most $k$ (of their primal graphs)
for which the space complexity of the equivalent {\sc nrobp}s  is at least $n^{ck}$. 

This result is a significant enhancement of the result of \cite{RazgonKR}.
Indeed, {\sc obdd}s are a subclass {\sc robp}s and  there is \emph{exponential separation} between the classes
(that is, there is a family of functions that can be represented by poly-size {\sc robp}s but require exponential size {\sc obdd}s).
A {\sc robp}, in turn, is a special case of ({\sc nrobp}) and there is an exponential separation between {\sc robp}s
and {\sc nrobp}s (\cite{WegBook}, Corollary 10.2.3). 
Thus the proposed result shows that read-once branching programs are inherently incapable to efficiently compute {\sc cnf} formulas of bounded
treewidth. 

We also demonstrate that the proposed result can be used in the non-parameterized context.
In particular, using this result, we provide a quasi-polynomial separation between {\sc nrobp}s
and a subset of decomposable negation normal forms ({\sc dnnf}s) \cite{DarwicheJACM} known as decision-{\sc dnnf}.
More precisely, we demonstrate a family of {\sc cnf} formulas that can be expressed as decision
{\sc dnnf}s of size $O(n^5)$ but the space complexity of {\sc nrobp}s is $n^{\Omega(\log n)}$.
The motivation for this result is described below. 

{\sc dnnf} is a representation of Boolean functions well known in the areas of knowledge representation
and databases. {\sc dnnf}s are much more succinct than {\sc robp}s. In fact a {\sc robp} 
can be seen as a special  case of {\sc dnnf} \cite{DerMar} and there is an exponential separation between 
these two representations. Like in the case of {\sc obdd}s, transformation from a {\sc cnf} formula to an
equivalent {\sc dnnf} is an important operation in the related applications. One remarkable property
of {\sc dnnf}s is their {\sc fpt} space complexity on {\sc cnf}s formulas with bounded treewidth. In particular,
a {\sc cnf} formula with treewidth $k$ can be transformed into a {\sc dnnf} of size $O(2^kn)$. In fact this
property is preserved for a number of restricted {\sc dnnf} subclasses, one of them is known as 
decision-{\sc dnnf} \cite{DesDNNF}. Interestingly, the possibility of exponential separation from {\sc robp} is not 
preserved for decision-{\sc dnnf}s: it has been shown in \cite{BeameDNNF} that a decision-{\sc dnnf} of size $N$
can be simulated by a {\sc robp} of size $O(N^{\log N})$. Our result shows that this upper bound is essentially
tight. Indeed, since {\sc robp} is a special case of {\sc nrobp}, this result implies
quasi-polynomial separation between {\sc robp} and decision-{\sc dnnf}, essentially matching the upper bound 
of \cite{BeameDNNF}.

We believe the proposed parameterized lower bound is interesting from the parameterized complexity theory perspective because
it contributes to the understanding of (concrete) parameterized \emph{space} complexity of various representations
of Boolean functions. We see at least two reasons why this research direction is worth to explore.
First, the results of of this kind are closely related (through substitution of the parameters with
appropriate functions of $n$) to the classical, non-parameterized complexity of Boolean function.
For example, the famous result of Razborov providing the first non-polynomial lower bound for the
space complexity of monotone circuits can be seen formulated in the parameterized setting as 
a space $n^{\sqrt{k}}$ lower bound for monotone circuits testing whether the given  graph has a clique of size $k$ \cite{Yukna}.

The second reason why we believe that the parameterized complexity of Boolean functions
is an interesting research direction is that parameterized upper bounds on the space
complexity of Boolean functions are important in applications related to verification, knowledge
representation, and databases. In fact, quite a few such upper bounds are already known 
(e.g. \cite{DarwicheJACM,SDD,DesDNNF,McMillan94,OBDDTWJha}).
Therefore, it is interesting to see if advanced parameterized complexity methodologies can be
applied in order to enhance these upper bounds and to obtain new ones.

\subsection{Overview of the proofs}
To prove the proposed parameterized lower bound, we use monotone $2$-{\sc cnf} formulas (their clauses are of form $(x_1 \vee x_2)$ where
$x_1$ and $x_2$ are $2$ distinct variables). These {\sc cnf} formulas are in one-to-one correspondence with graphs
having no isolated vertices: variables correspond to vertices and $2$ variables occur in the same clause if and only if the
corresponding vertices are adjacent. This correspondence allows us to use these {\sc cnf} formulas and graphs interchangeably.
We introduce the notion of Matching Width ({\sc mw}) of a graph $G$ and prove two theorems.
One of them (Theorem \ref{nrobplbdmw}) states that
a {\sc nrobp} equivalent to a monotone 2-{\sc cnf} formula with the corresponding graph $G$ having {\sc mw} at least $t$ is of size at least $2^{t/a}$ where $a$
is a constant dependent on the max-degree of $G$. The second theorem (Theorem \ref{dmwtw}) states that for each sufficiently large $k$ there is an
infinite family of graphs of treewidth $k$ and max-degree $5$ whose {\sc mw} is at least $\log n*k/b$ for some constant $b$ independent of $k$. 
The main theorem immediately follows from replacement of $t$ in the former lower bound by the latter one.

The proof of Theorem \ref{nrobplbdmw} uses the following combinatorial statement.
Let ${\bf VC}(G)$ be the set of all vertex covers of a graph $G$ and let ${\bf S}$ be a family
of subsets of $V(G)$ of size at least $t$ such that each element of ${\bf VC}(G)$ is a superset of
some element of ${\bf S}$. Then $|{\bf S}| \geq 2^{t/a}$ where $a$ is a universal constant 
as in the previous paragraph. 

In order to define the family of graphs for Theorem \ref{dmwtw}, we introduce
graphs $T_r(H)$ where $T_r$ is a complete binary tree of height $r$ and $H$ is an arbitrary graph.
In the graph $T_r(H)$ each vertex of $T_r$ is replaced by a copy of $H$. Copies corresponding to adjacent
vertices of $T_r$ are connected by edges so that each vertex of $H$ is connected to the `same' vertex of $H$ of
the adjacent copy. For the proof of Theorem \ref{dmwtw}, we take as $H$ a path of length about $k/2$.

The strategy outlined above is similar to that we used in \cite{RazgonKR}. However, there are two essential differences.
First, due to a much more `elusive' nature of {\sc nrobp}s compared to that of {\sc obdd}, the counting argument is more
sophisticated and more restrictive: it applies only to {\sc cnf} formulas whose graphs are of constant degree. Due to this latter
aspect, the family of graphs requires a more delicate construction and reasoning.

The rest of the paper is organized as follows. Section 2 introduces the necessary background.
Section 3,4, and 5 prove the parameterized lower bound (the last two sections prove auxiliary
theorems used for the lower bound proof in section 3). Section 6 establishes the quasipolynomial
separation  between decision{\sc dnnf} and {\sc nrobp}. Finally, two sections in the Appendix 
demonstrate validity of our assumptions regarding {\sc nrobp}, see Section 2 for further details.

\section{Preliminaries}
In this paper when we refer to a  \emph{set of literals} we assume that it does not
contain an occurrence of a variable and its negation.
For a set $S$ of literals we denote by $Var(S)$ the set of variables
whose literals occur in $S$. If $F$ is a Boolean function
or its representation by a specified structure, we denote by $Var(F)$
the set of variables of $F$. A truth assignment to $Var(F)$ on which $F$
is true is called a \emph{satisfying assignment} of $F$. A set $S$ of literals
represents the truth assignment to $Var(S)$ where variables occurring
positively in $S$ (i.e. whose literals in $S$ are positive) are assigned with $true$
and the variables occurring negatively are assigned with $false$.
We denote by $F_S$ a function whose set of satisfying assignments consists of all sets $S'$ of literals
such that $S \cup S'$ is a satisfying assignment of $F$. We call $F_S$ a \emph{subfunction}
of $F$. 



\begin{definition} \label{arosrn}
A non-deterministic read-once branching program ({\sc nrobp})
$Y$ implementing (computing) a function $F$ is a directed acyclic graph ({\sc dag}) (with possible multiple edges)
with one leaf, one root, and with some edges labelled by literals of the variables of $F$ in a way that 
there is no directed path having two edges labelled with literals of the same variable. We denote by $A(P)$ the set of literals
labelling edges of a directed path $P$ of $Y$.

The connection between $Y$ and $F$ is defined as follows.
Let $P$ be a path from the root to the leaf of $Y$. Then any set of literals $A \supseteq A(P)$  
such that $Var(A)=Var(F)$ is a satisfying assignment of $F$.
Conversely, let $A$ be a satisfying assignment of $F$. Then there is a path $P$ from the root to
the leaf of $Y$ such that $A(P) \subseteq A$.
\end{definition}

{\bf Remark.}
A traditional definition of a {\sc nrobp} is a {\sc robp} with guessing nodes. 
Definition \ref{arosrn} in fact introduces acyclic read-once switching and rectifier networks ({\sc arosrn}s).
However, these models are equivalent in the sense that an {\sc arosrn} can simulate {\sc nrobp} without
increase of the number of edges and a {\sc nrobp} can simulate an {\sc arosrn} with at most three times increase
of the number of edges. The details are provided in Appendix B. 
The equivalence of these models is mentioned in \cite{JuknaECCC94}.

We say that a {\sc nrobp} $Y$ is \emph{uniform} if the following is true.
Let $a$ be a node of $Y$ and let $P_1$ and $P_2$ be $2$ paths from the root
of $Y$ to $a$. Then $Var(A(P_1))=Var((A(P_2))$. That is, these paths are labelled
by literals of the same set of variables. Also, if $P$ is a path from the root to the leaf of $Y$
then $Var(A(P))=Var(F)$. Thus there is a one-to-one correspondence between the sets of literals labelling
paths from the root to the leaf of $Y$ and the satisfying assignments of $F$.

{\bf All the {\sc nrobp}s considered in Sections 3-6 of this paper are uniform.}
This assumption does not affect our main result because
an arbitrary {\sc nrobp} can be transformed into a uniform one at the price of $O(n)$ times increase of
the number of edges. 
For the sake of completeness, we provide the transformation and its correctness
proof in Appendix A. We use the construction described in the proof sketch of Proposition 2.1 of \cite{RaWiYao}.

Now we are going to define the Decomposable Negation Normal Form ({\sc dnnf}) and its subclass decision-{\sc dnnf}
for which we prove a separation result in Section 6.

{\bf Remark.} The only thing we need to know for this separation result is that a {\sc cnf} formula with a bounded 
primal graph treewidth can be transformed into an {\sc fpt}-size decision-{\sc dnnf} \cite{DesDNNF}. That is, the two paragraphs 
below are not needed for the technical reasoning. We provide these definitions for the sake of completeness in the 
sense that all the representations of Boolean functions occurring in the statements of this paper are explicitly
defined. 

Recall that a Boolean circuit over the $\vee,\wedge,\neg$ is called \emph{de Morgan} circuit if the negations
are applied only to the input (variable) gates. Next, we define a \emph{decomposable node}. Let $x$ be a gate
of a Boolean circuit $X$. We denote by $VReach(x)$ the set of variables such that $x$ is reachable from their
respective input gates. We say that $x$ is \emph{decomposable} if for any two in-neighbours $y_1$ and $y_2$
of $x$, $VReach(y_1) \cap VReach(y_2)=\emptyset$. A {\sc dnnf} is a de-Morgan circuit with all the {\sc and}-nodes
being decomposable. 

\begin{figure}[h]
\includegraphics[height=4cm]{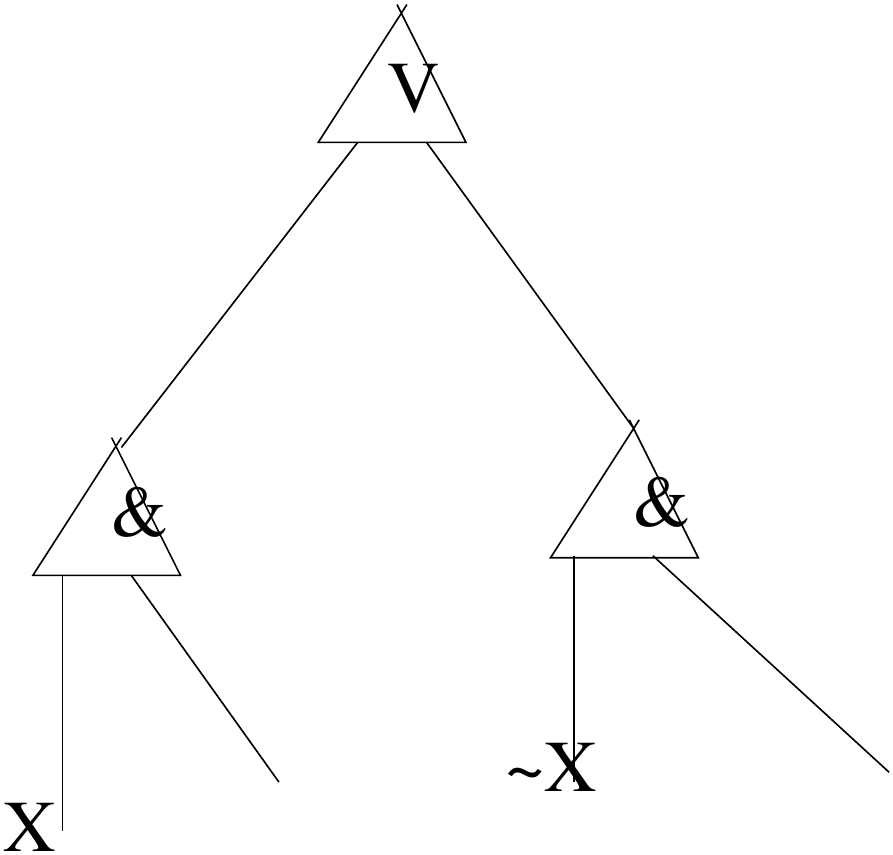}
\caption{A decision node}
\label{desnode}
\end{figure}

\begin{figure}[h]
\includegraphics[height=5cm]{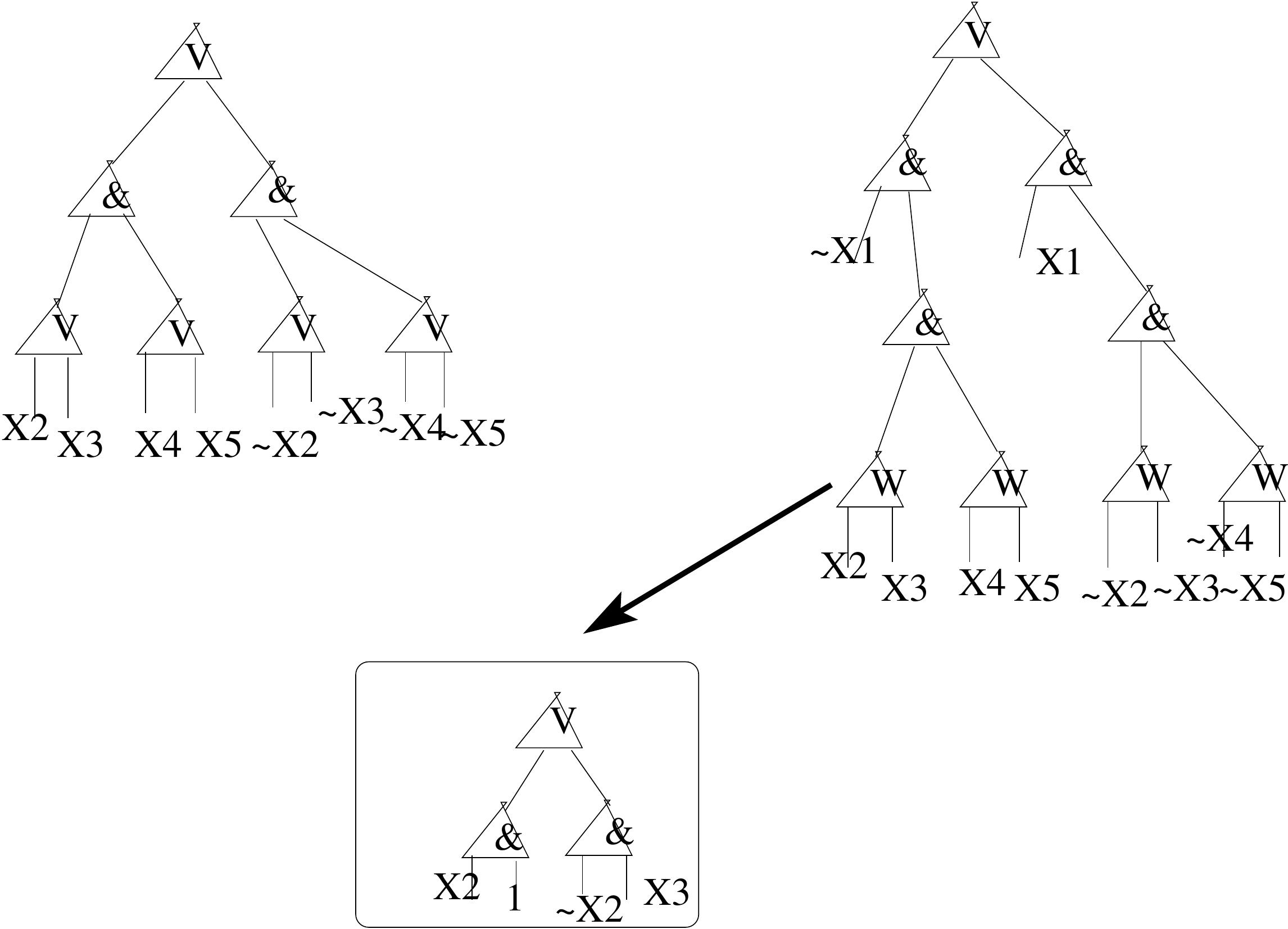}
\caption{A {\sc dnnf} and a decision-{\sc dnnf} for a function 
         $(x_1 \vee x_2 \vee x_3)(x_1 \vee x_4 \vee x_5)(\neg x_1 \vee \neg x_2 \vee \neg x_3)(\neg x_1 \vee \neg x_4 \vee \neg x_5)$.
For the sake of compactness, we introduced a `macro-node' $W$ that expresses a disjunction of two literals in terms of decision-{\sc dnnf}.}
\label{dnnfs}
\end{figure}

We say that an {\sc or}-node $x$ of a {\sc dnnf} is a decision node (see Figure \ref{desnode}) if it is binary, both its in-neighbours $y_1$ and $y_2$ are {\sc and}-nodes
and there is a variable $x$ such that $x$ is an input of, say $y_1$ and $\neg x$ is an input of $y_2$. 
A {\sc dnnf} is called \emph{decision} {\sc dnnf} if all its {\sc or} nodes are decision ones. See Figure \ref{dnnfs} showing a {\sc dnnf}
and a decision-{\sc dnnf} for the same function. Note that for the latter we use both variable and constant input gates. 

Given a graph $G$, its \emph{tree decomposition} is a pair $(T,{\bf B})$ where $T$ 
is a tree and ${\bf B}$ is a set of bags $B(t)$ corresponding to the vertices $t$ of $T$.
Each $B(t)$ is a subset of $V(G)$ and the bags obey the rules of \emph{union} (that is, $\bigcup_{t \in V(T)} B(t)=V(G)$),
\emph{containment} (that is, for each $\{u,v\} \in E(G)$ there is $t \in V(t)$ such that $\{u,v\} \subseteq B(t)$),
and \emph{connectedness} (that is for each $u \in V(G)$, the set of all $t$ such that $u \in B(t)$ induces a subtree of $T$).
The \emph{width} of $(T,{\bf B})$ is the size of the largest bag minus one. The treewidth of $G$ is the smallest width of a tree
decomposition of $G$.

Given a {\sc cnf} formula $\phi$, its \emph{primal graph} has the set of vertices corresponding to the variables of $\phi$.
Two vertices are adjacent if and only if there is a clause of $\phi$ where the
corresponding variables both occur.

\section{The parameterized lower bound}
\begin{figure}[h]
\includegraphics[height=5cm]{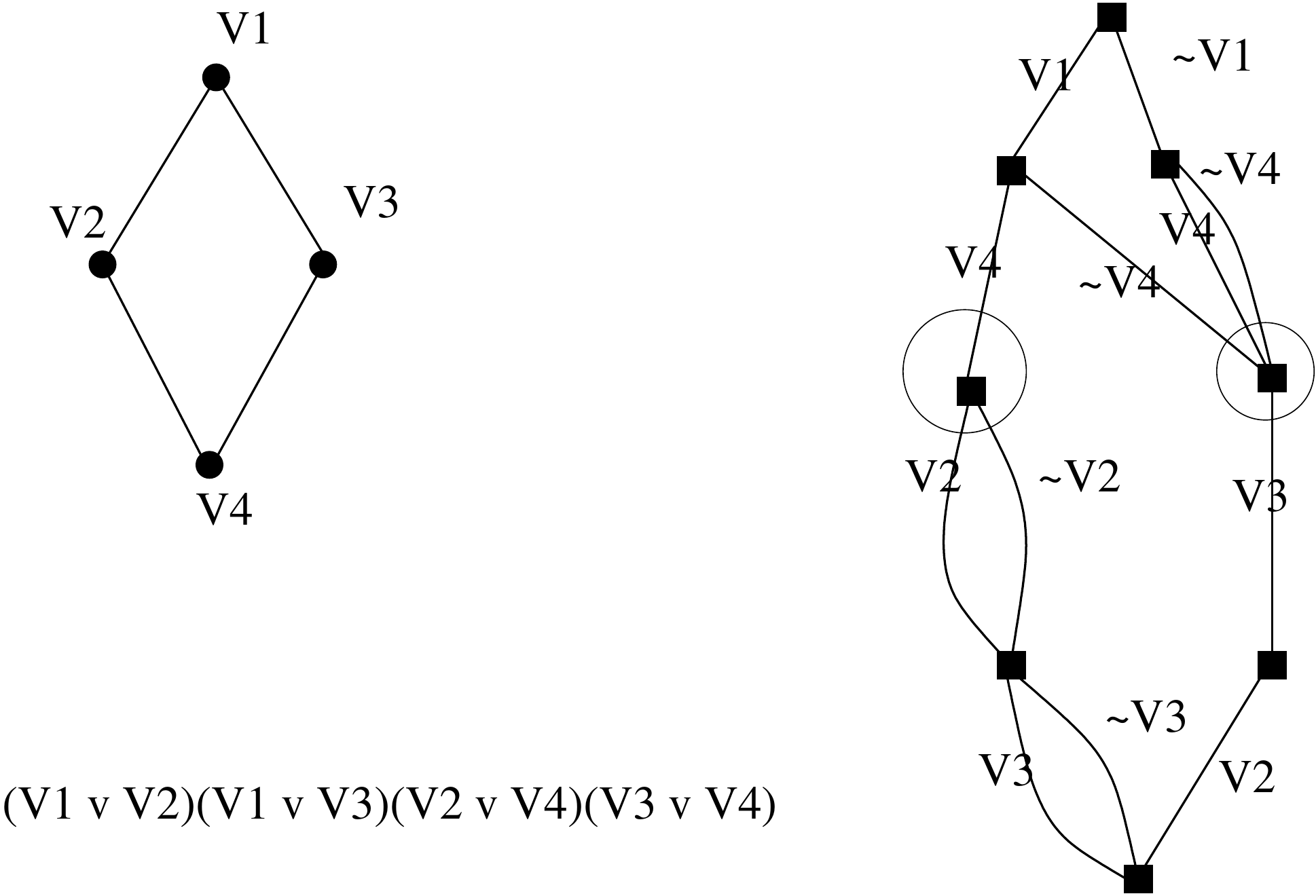}
\caption{A graph, the corresponding {\sc cnf} formula and a {\sc nrobp} of the {\sc cnf} formula. Circles denote the nodes
used in Section 4 for illustration of the definition of a $t$-node.}
\label{srn}
\end{figure}

A \emph{monotone} 2-{\sc cnf} formula has clauses
of the form $(x \vee y)$ where $x$ and $y$ are two distinct variables. 
Such {\sc cnf} formulas can be put in one-to-one correspondence
with graphs that do not have isolated vertices. In particular, let $G$ be such
a graph. Then $G$ corresponds to a 2{\sc cnf} formula $\phi(G)$ whose variables are the
vertices of $G$ and the set of clauses is $\{(u \vee v)|\{u,v\} \in E(G)\}$.
These notions, together with the corresponding {\sc nrobp}, are illustrated on
Figure \ref{srn}. \footnote{Notice that on the {\sc nrobp} in Figure \ref{srn}, there is
a path where $v_2$ occurs before $v_3$ and a path where $v_3$ occurs before $v_2$.
Thus this {\sc nrobp}, although uniform, is not \emph{oblivious}.}
It is not hard to see that $G$ is the primal graph of $\phi(G)$, hence we can refer to the 
treewidth of $G$ as the primal graph treewidth of $\phi(G)$.

The following theorem is the main result of this paper. 

\begin{theorem} \label{maintheor}
There is a constant $c$ such that
for each $k \geq 3$ there is an infinite class ${\bf G}$ of graphs each
of treewidth of at most $k$ such that for each $G \in {\bf G}$,
the smallest {\sc nrobp} equivalent to $\phi(G)$ is of size at least ${n}^{k/c}$,
where $n$ is the number of variables of $\phi(G)$.
\end{theorem}

In order to prove Theorem \ref{maintheor}, we 
introduce the notion of \emph{matching width} ({\sc mw})
of a graph and state two theorems proved in the subsequent two sections.
One claims that if the max-degree of $G$ is bounded then the size of a {\sc nrobp} realizing $\phi(G)$
is exponential in the {\sc mw} of $G$. The other theorem claims that for each sufficiently large $k$ 
there is an infinite class of graphs of bounded degree and of treewidth at most $k$ whose {\sc mw} 
is at least $\log n*k/b$ for some universal constant $b$. 
Theorem \ref{maintheor} will follow as an immediate corollary of these two theorems. 

\begin{definition}{\bf Matching width.}\\
Let $SV$ be a \emph{permutation} of $V(G)$ 
and let $S_1$ be a \emph{prefix} of $SV$ (i.e. all vertices of $SV \setminus S_1$ are ordered after
$S_1$). The \emph{matching width} of $S_1$ is the size of the largest matching
consisting of the edges between $S_1$ and $V(G) \setminus S_1$.
\footnote{We sometimes treat sequences as sets, the correct use will be always clear
from the context} The \emph{matching width} of $SV$ is the largest \emph{matching
width} of a prefix of $SV$. The \emph{matching width} of $G$, denoted by $mw(G)$, is the smallest 
matching width of a permutation of $V(G)$. 
\end{definition}

\paragraph{Remark.} The above definition of matching width is a special 
case of the notion of \emph{maximum matching width} as defined
in \cite{VaThesis}. 

To illustrate the notion of matching width recall that $C_n$ and $K_n$ respectively
denote a cycle and a complete graph of $n$ vertices. Then, for a sufficiently large $n$,
$mw(C_n)=2$. On the other hand $mw(K_n)=\lfloor n/2 \rfloor$. 

\begin{theorem} \label{nrobplbdmw}
There is a function $f$ such that
for any graph $G$ the size of
{\sc nrobp} realizing $\phi(G)$ is at least $2^{mw(G)/f(x)}$
where $x$ is the max-degree of $G$. 
\end{theorem}


\begin{theorem} \label{dmwtw}
There is a constant $b$ such that
for each $k \geq 3$ there is an infinite class ${\bf G}$ of graphs of degree at most $5$ 
such that the treewidth of all the graphs of $G$ is at most $k$ and the matching width
of each $G \in {\bf G}$ is at least $(log n*k)/b$ where $n=|V(G)|$.
\end{theorem}


Now we are ready to prove Theorem \ref{maintheor}.

{\bf Proof of Theorem \ref{maintheor}.}
Let ${\bf G}$ be the class whose existence is claimed by Theorem \ref{dmwtw}.
By Theorem \ref{nrobplbdmw}, for each $G \in {\bf G}$ the size of a {\sc nrobp} 
realizing $\phi(G)$ is of size at least $2^{mw(G)/f(5)}$.
Further on, by Theorem \ref{dmwtw}, $mw(G) \geq (\log n*k)/b$, for some constant $b$.
Substituting the inequality for $mw(G)$ into the lower bound $2^{mw(G)/f(5)}$
supplied by Theorem \ref{nrobplbdmw}, we get that the size
of a {\sc nrobp} is at least $2^{\log n *k/c}$ where $c=f(5)*b$. Replacing $2^{\log n}$ by $n$
gives us the desired lower bound. $\blacksquare$

From now on, the proof is split into two independent parts: Section \ref{lbmainproof} proves
Theorem \ref{nrobplbdmw} and Section \ref{dmwmainproof} proves Theorem \ref{dmwtw}.

\section{Proof of Theorem \ref{nrobplbdmw}} \label{lbmainproof}
Recall that we are going to prove that for any graph $G$, the size of 
a {\sc nrobp} computing $\phi(G)$ is at least $2^{mw(G)/f(x)}$ where $f(x)$ is
a universal function depending on the max-degree $x$ of $G$ only. 

Recall that the vertices of graph $G$ serve as variables in $\phi(G)$.
That is, in the truth assignments to $Var(\phi(G))$, the vertices are treated
as literals and may occur positively or negatively. Similarly for a path $P$ of a
{\sc nrobp} $Z$ implementing $\phi(G)$, we say that a vertex $v \in V(G)$ \emph{occurs}
on $P$ if either $v$ and $\neg v$ labels an edge of $P$.
In the former case this is a \emph{positive occurrence}, in the latter case a \emph{negative} one.

Recall that a Vertex Cover ({\sc vc}) of $G$ is $V' \subseteq V(G)$ incident to all
the edges of $E(G)$. 

\begin{observation} \label{basicobs}
$S$ is a satisfying assignment of $\phi(G)$ if and only if the vertices of $G$ 
occurring positively in $S$ form a {\sc vc} of $G$. Equivalently, $V' \subseteq V(G)$
is the set of all vertices of $G$ occurring positively on a root-leaf path of $Z$ if
and only if $V'$ is a {\sc vc} of $G$.
\end{observation}

In light of Observation \ref{basicobs}, we denote the set of all vertices occurring positively
on a root-leaf path $P$ of $Z$ by $VC(P)$.

The proof of Theorem \ref{nrobplbdmw} requires two intermediate statements. 
For the first statement, let $a$ be a node of an {\sc nrobp} $Z$. 
For an integer $t>0$, we call $a$ a \emph{$t$-node} if there is a set $S(a)$
of size at least $t$ such that for each root-leaf path $P$ passing through $a$,
$S(a) \subseteq VC(P)$. To demonstrate the notion of a $t$-node, consider the two nodes
denoted by circles in Figure \ref{srn}. They are $2$-nodes for the given {\sc nrobp},
the witnessing set for the left-hand node is $\{v_1,v_4\}$ and for the right-hand node
is $\{v_2,v_3\}$.

\begin{lemma} \label{tnodecut}
Suppose that the matching width of $G$ is at least $t$.
Then any root-leaf path of $Z$ contains a $t$-node or, put it differently,
$t$-nodes of $Z$ form a root-leaf cut.
\end{lemma}

{\bf Proof.}
We need to show that each root-leaf path $P$ passes through a $t$-node.
Due to the uniformity of $Z$, (the vertices of $G$ corresponding to) the labels of $P$ being explored from the root
to the leaf form a permutation $SV$ of $V(G)$. Let $SV'$ be a prefix of the permutation
witnessing the matching width at least $t$. In other words, there is a matching 
$M=\{\{u_1,v_1\}, \dots, \{u_t,v_t\}\}$ of $G$ such that all of $u_1, \dots, u_t$ belong to $SV'$,
while all of $v_1, \dots, v_t$ belong to $SV \setminus SV'$.
Let $u$ be the last vertex of $SV'$ and let $a$ be the head of the edge of $P$ whose label is a literal of $u$.
We claim that $a$ is a $t$-node with a witnessing set $S(a)=\{x_1, \dots, x_t\}$ such that $x_i \in \{u_i,v_i\}$
for each $x_i$.

Indeed, observe that for each $\{u_i,v_i\}$ there is $x_i \in \{u_i,v_i\}$ such that $x_i \in VC(P)$
for each root-leaf path $P$ passing through $a$. Clearly for any root-leaf path $Q$ of $Z$, either $u_i \in VC(Q)$
or $v_i \in VC(Q)$ for otherwise $VC(Q)$ is not a {\sc vc} of $G$ in contradiction to Observation \ref{basicobs}. 
Thus if such $x_i$ does not exist then 
there are two paths $Q^1$ and $Q^2$ meeting $a$ such that 
$VC(Q^1) \cap \{u_i,v_i\}=\{u_i\}$ and $VC(Q^2) \cap \{u_i,v_i\}=\{v_i\}$.

For a root-leaf path $Q$ passing through $a$ denote by $Q_a$ the prefix of $Q$ ending with $a$ and by
$\neg Q_a$ the suffix of $Q$ beginning with $a$. Note that by definition of $SV'$, $u_i$
occurs in $P_a$ and $v_i$ occurs in $\neg P_a$. 
By uniformity of $Z$, $Var(P_a)=Var(Q^1_a)=Var(Q^2_a)$ and hence
it follows that $u_i$ occurs both in $Q^1_a$ and $Q^2_a$.
Similarly we establish that $v_i$ occurs in both $\neg Q^1_a$ and $\neg Q^2_a$.
It remains to observe that, by definition, $u_i$ occurs negatively in $Q^2_a$
and $v_i$ occurs negatively in $\neg Q^1_a$. Hence $Q^*=Q^2_a+\neg Q^1_a$ is a root-leaf
path of $Z$ such that $VC(Q^*)$ is disjoint with $\{u_i,v_i\}$, a contradiction to Observation \ref{basicobs}, 
confirming the existence of the desired $x_i$. 

Suppose that there is a root-leaf path $P'$ of $Z$ passing through $a$ such that $S(a) \nsubseteq VC(P')$.
This means that there is $x_i \notin VC(P')$ contradicting the previous two paragraphs.
Thus being $a$ a $t$-node has been established and the lemma follows. $\blacksquare$

For the second statement, let ${\bf A}$ and ${\bf B}$ be two families of subsets of a universe ${\bf U}$.
We say that ${\bf A}$ \emph{covers} ${\bf B}$ if for each $S \in {\bf B}$ there is
$S' \in {\bf A}$ such that $S' \subseteq S$. If each element of ${\bf A}$ is of size at least $t$
then we say that ${\bf A}$ is a $t$-\emph{cover} of ${\bf B}$.
Denote by ${\bf VC}(G)$ the set of all VCs of $G$.

\begin{theorem} \label{lbengine}
There is a function $f$ such that the following is true.
Let $H$ be a graph.
Let ${\bf A}$ be a $t$-cover of ${\bf VC}(H)$. 
The $|{\bf A}| \geq 2^{t/f(x)}$ where $x$ is the max-degree of $H$.
\end{theorem}

The proof of Theorem \ref{lbengine}, using a probabilistic argument, 
is provided in Subsection \ref{lbenproof}.
See \cite{Bova14} (Theorem 3 and Corollary 2) for a non-probabilistic proof.

Now we are ready to prove Theorem \ref{nrobplbdmw}.

{\bf Proof of Theorem \ref{nrobplbdmw}.}
Let $N$ be the set of all $t$-nodes of $Z$.
For each $a \in N$, specify one $S(a)$ of size at
least $t$ such that for all paths $P$ of $Z$ passing
through $a$, $S(a) \subseteq VC(P)$. 
Let ${\bf S}=\{S_1, \dots, S_q\}$ be the set of all such $S(a)$.
Then we can specify \emph{distinct} $a_1, \dots, a_q$
such that $S_i=S(a_i)$ for all $i \in \{1, \dots, q\}$.

Observe that ${\bf S}$ covers ${\bf VC}(G)$.
Indeed, let $V' \in {\bf VC}(G)$. By Observation \ref{basicobs}, there is a
root-leaf path $P$ with $V'=VC(P)$. By Lemma \ref{tnodecut}, $P$ passes through some
$a \in N$ and hence $S(a) \subseteq VC(P)$. By definition, $S(a)=S_i$
for $i \in \{1, \dots, q\}$ and hence $S_i \subseteq V'$.
Thus ${\bf S}$ is a $t$-cover of ${\bf VC}(G)$.

It follows from Theorem \ref{lbengine} that $q=|{\bf S}| \geq 2^{t/f(x)}$
where $x$ is a max-degree of $G$ and $f$ is a universal function independent on
$G$ or $t$. It follows that $Z$ contains at least $2^{t/f(x)}$ distinct
nodes namely $a_1, \dots, a_q$. $\blacksquare$

\subsection{Proof of Theorem \ref{lbengine}} \label{lbenproof}
Denote $E(H)$ by $E$. 
For each $e=\{u,v\}$, we toss a fair coin whose outcomes are $u$ or $v$
and the denote the outcome by $Out(e)$. 
For $E' \subseteq E$, let $Out(E')=\{Out(e)|e \in E'\}$. 
That is, $Out(E')$ is a random set consisting of ends of edges of $E'$ each
chosen uniformly at random.

\begin{claim}
Let $S \subseteq V(G)$.
Then $Pr(S \subseteq Out(E)) \leq (1-2^{-x})^{|S|/(x+1)}$
\end{claim}

Let us see how the claim implies the statement of the lemma.
Let ${\bf A}$ be as in the statement of the lemma. 
Then, by the claim above and the union bound, the probability
that at least one element of ${\bf A}$ is a subset of $Out(E)$
is at most $|{\bf A}|*(1-2^{-x})^{t/(x+1)}=|{\bf A}|*2^{-t/f(x)}$
where $f$ is a function such that $2^{-1/f(x)}=(1-2^{-x})^{1/(x+1)}$. 
Suppose that $|{\bf A}|<2^{t/f(x)}$.
Then the above probability is smaller than $1$. That is, there is
a set $T$ obtained by choosing one end of each $e \in E$ such that
$T$ is not a superset of any element of ${\bf A}$. By construction
$T$ is a VC of $H$. Thus we have just observed that any family of 
less than $2^{t/f(x)}$ subsets of $V(H)$ of size at least $t$ cannot
cover all of ${\bf VC}(H)$, as required.

{\bf Proof of the claim.}
For $u \in V(H)$, denote by $E_u$ the set of edges incident to $u$. 
For $S \subseteq V(H)$, let $E_S=\bigcup_{u \in S} E_u$. 
Then it is easy to notice the following.
\begin{equation} \label{eq1}
u \in Out(E) \Leftrightarrow u \in Out(E_u)
\end{equation}  

\begin{equation} \label{eq2}
Pr(u \in Out(E))=Pr(u \in Out(E_u))
\end{equation}

Furthermore, for a set $S$,

\begin{equation} \label{eq3}
S \subseteq Out(E) \Leftrightarrow S \subseteq Out(E_S)
\end{equation}

We will also need the following form of statement that the event $u \in Out(E)$
is independent on the guessed ends of edges outside $E_u$.
In particular, let $E' \subseteq E$ be such that $E_u \cap E'=\emptyset$
and let $S \subseteq V(H)$ be such that $Pr(S \subseteq Out(E'))>0$. Then
\begin{equation} \label{eq4}
Pr(u \in Out(E)|S \subseteq Out(E'))=Pr(u \in Out(E_u)|S \subseteq Out(E'))=Pr(u \in Out(E_u))
\end{equation}

Let $I=\{u_1, \dots, u_q\}$ be an independent set of $H$.
Then the sets $E_{u_i}$ are pairwise disjoint and, in particular, for
each $i<q$, $E^i=\bigcup_{j \leq i} E_{u_i}$ is disjoint with $E_{u_{i+1}}$.
We prove by induction on $q$ that $Pr(I \subseteq Out(E))=\prod_{i=1}^q Pr(u_i \in Out(E_{u_i}))$.
For $q=1$ the claim immediately follows from (\ref{eq2}).
Assume that $q>1$.
Then
\begin{equation} \label{eq5}
Pr(I \subseteq Out(E))=Pr(I \setminus \{u_q\} \subseteq Out(E))*Pr(u_q \in Out(E)|I \setminus \{u_q\} \subseteq Out(E))
\end{equation} 

By the induction assumption,
\begin{equation} \label{eq6}
Pr(I \setminus \{u_q\} \subseteq Out(E))=\prod_{i=1}^{q-1} Pr(u_i \in Out(E_{u_i}))
\end{equation}

Also,
\begin{equation} \label{eq7}
\begin{aligned} 
&Pr(u_q \in Out(E)|I \setminus \{u_q\} \subseteq Out(E))=Pr(u_q \in Out(E)|I \setminus \{u_q\} \subseteq Out(E_{I \setminus q}))\\
&=Pr(u_q \in Out(E_{u_q}))
\end{aligned}
\end{equation}

the first equality follows from (\ref{eq3}), the second from (\ref{eq4}).
Replacing the factors of the right part of (\ref{eq5}) with the respective right parts of (\ref{eq6})
and (\ref{eq7}), we obtain 
$Pr(I \subseteq Out(E))=[\prod_{i=1}^{q-1} Pr(u_i \in Out(E_{u_i}))]*Pr(u_q \in Out(E_{u_q}))$
as required. 

Notice further that $Pr(u \in Out(E_u))=1-2^{-|E_u|} \leq 1-2^{-x}$.
Hence, $Pr(I \subseteq Out(E)) \leq (1-2^{-x})^{|I|}$.
Now, consider an arbitrary $S \subseteq V(H)$. Then there is an independent set
$I \subseteq S$ of size at least $|S|/(x+1)$ (recall that $x$ is the max-degree of $H$).
Hence $Pr(S \subseteq Out(E)) \leq Pr(I \subseteq Out(E)) \leq (1-2^{-x})^{|S|/(x+1)}$
as required.

\section{Proof of Theorem \ref{dmwtw}} \label{dmwmainproof}
Recall that we are going to prove that for each $k \geq 3$ 
there is an infinite class of graphs of degree at most $5$ having treewidth $k$ and matching width
at least $(\log n*k)/b$ where $b$ is a universal constant. 

\begin{figure}[h]
\includegraphics[height=2.5cm]{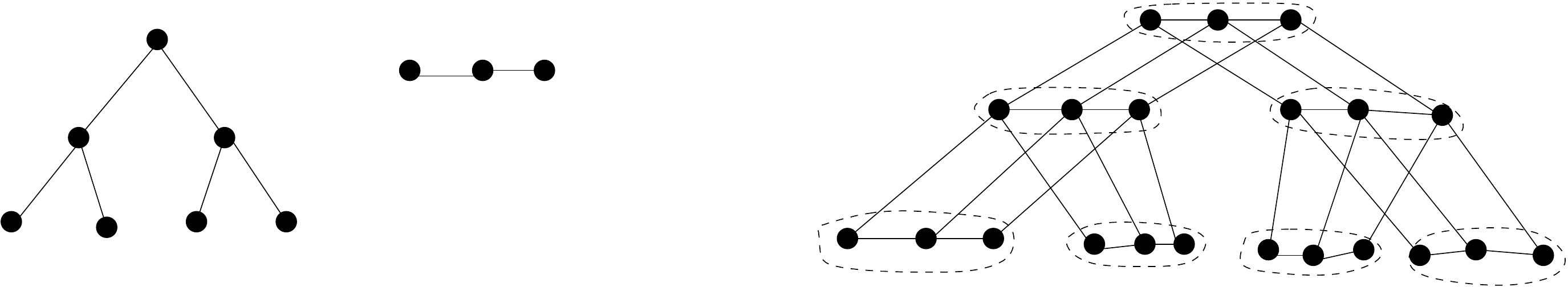}
\caption{Graphs from the left to the right: $T_3,P_3,T_3(P_3)$. The dotted ovals surround the copies of $P_3$ in $T_3(P_3)$.}
\label{trees}
\end{figure}

Let us define first a more general class of graphs for which the class of graphs used for the
proof of Theorem \ref{dmwtw} will be a subclass. 
Denote by $T_r$ a complete binary tree of height (root-leaf distance) $r$.
Let $T$ be a tree and $H$ be an arbitrary graph. 
Then $T(H)$ is a graph having disjoint copies of $H$ in one-to-one correspondence with the vertices of $T$. 
For each pair $t_1,t_2$ of adjacent vertices of $T$, the corresponding copies are
connected by making adjacent the pairs of \emph{same} vertices of these copies.
Put differently, we can consider $H$ as a labelled graph where all vertices are associated with distinct
labels. Then for each edge $\{t_1,t_2\}$ of $T$, edges are introduced between the vertices of the corresponding copies having
the same label. An example of this construction is shown on Figure \ref{trees}.

In order to prove Theorem \ref{dmwtw} we will consider all graphs $T_r(P_q)$ where $r$ gets ranges over all natural numbers
and $q$ is about $k/2$, the precise definition is provided below inside the proof. 
We also need to prove three structural lemmas about graphs $T(H)$, the first one being
an auxiliary statement for the second one and the second one being an auxiliary statement for the third one. 
Note that these structural lemmas do not restrict the structure of $H$, besides Lemma \ref{dmwtwstruct}
requiring $H$ to be connected. 

\begin{lemma} \label{matchontheway}
Suppose the vertices of $T(H)$ are partitioned into two subsets. 
Let $L$ be a subset of vertices of $H$ such that $|L|=t$.
Suppose there are two copies $H_1$ and $H_2$ of $H$ such that for each
$u \in L$ the copies of vertex $u$ in $H_1$ and $H_2$
belong to distinct partition classes. 
Then $T(H)$ has a matching of size $t$ with the ends of each edge lying
in different partition classes.
\end{lemma}

{\bf Proof.}
Let $v_1$ and $v_2$ be the respective vertices of $T$ 
corresponding to $H_1$ and $H_2$. Let $p$ be the path between $v_1$
and $v_2$ in $T$. Then for each $u \in L$ there are two consecutive
vertices $v'_1$ and $v'_2$ of this path with respective copies $H'_1$
and $H'_2$ such that the copy $u'_1$ of $u$ in $H'_1$ 
belongs to the same partition class as the copy $u_1$ of $u$ in $H_1$ 
and the copy $u'_2$ of $u$ in $H'_2$ belongs 
to the same partition class as the copy $u_2$  of $u$ in $H_2$. 
By construction, $T(H)$ has an edge $\{u'_1,u'_2\}$ which we choose to correspond to $u$.
Let $L=\{u^1,\dots u^t\}$ and consider the set of edges as above corresponding to each $u^i$. 
By construction, both ends of the edge corresponding to each $u^i$ are copies
of $u^i$ and also these ends correspond to distinct partition classes. 
It follows that these edges do not have joint ends and indeed 
constitute a desired matching of size $t$ $\blacksquare$

\begin{lemma} \label{mincase}
Let $T$ be a tree consisting of at least $p$ vertices. 
Let $H$ be a connected graph of at least $2p$ vertices.
Let $V_1,V_2$ be a partition of $V(T(H))$ such that both
partition classes contain at least $p^2$ vertices. Then 
$T(H)$ has a matching of size $p$ with the ends of each 
edge belong to distinct partition classes. 
\end{lemma}

{\bf Proof.}
Assume first that there are at least $p$ copies of $H$ corresponding
to vertices of $T$ that contain vertices of both partition classes. 
Since $H$ is a connected graph, for each copy we can specify an edge 
with one end in $V_1$ and the other end in $V_2$. These edges belong 
to disjoint copies of $H$, hence none of these edges have a common end. 
Since there are $p$ copies of $H$, we have the desired matching of size $p$.

If the assumption in the previous paragraph is not true then, since $T$
has at least $p$ vertices, there is a vertex $u$ of $T$ such that the copy 
$H_1$ of $H$ corresponding to $u$ contains vertices of only one partition 
class; assume w.l.o.g. that this class is $V_1$. We call $u$ a \emph{non-partitioned vertex} of $T$.
Then there is a vertex $v$ of $T$ such that the copy $H_2$ of $H$ corresponding 
to $v$ contains at least $p$ vertices of $V_2$.
Indeed, otherwise, the vertices of the copies of $H$ associated with the non-partitioned
vertices of $T$ all belong to $V_1$. Consequently, vertices of $V_2$ can occur only
in the remaining at most $p-1$ copies of $H$. If each of these copies contains at most $p-1$ 
vertices of $V_2$ then the total number of vertices of $V_2$ is smaller than $p^2$ in
contradiction to our assumption. We conclude that the required vertex $v$ indeed exists.

Let $L$ be the set of vertices of $H$ 
whose copies in $H_2$ belong to $V_2$. By assumption, all the copies of $L$ 
in $H_1$ belong to $V_1$. By Lemma \ref{matchontheway},
$H_1$ and $H_2$ witness the existence of a matching of size $p$ with ends of each
edge belonging to distinct partition classes. $\blacksquare$ 

\begin{lemma} \label{dmwtwstruct}
Let $p$ be an arbitrary integer and let $H$ be an arbitrary connected graph of $2p$ vertices.
Then for any $r \geq \lceil log p \rceil$, $mw(T_r(H)) \geq (r+1-\lceil log p \rceil)p/2$.
\end{lemma}

{\bf Proof.}
The proof is by induction on $r$.
The first considered value of $r$ is $\lceil log p \rceil$.
After that $r$ will increment in $2$. In particular,
for all values of $r$ of the form $\lceil log p \rceil+2x$,
we will prove that $mw(T_r(H)) \geq (x+1)p$ and, moreover, for
each permutation $SV$ of $V(T_r(H))$, the required matching can be witnessed
by a partition of $SV$ into a suffix and a prefix of size at least $p^2$ each.
Let us verify that the lower bound $mw(T_r(H)) \geq (x+1)p$ implies the lemma.
Suppose that $r=\lceil log p \rceil+2x$ for some non-negative integer $x$.
Then $mw(G) \geq (x+1)p=((r-\lceil log p \rceil)/2+1)p>(r-\lceil log p \rceil+1)p/2$. Suppose $r=\lceil log p \rceil+2x+1$.
Then $mw(G)=mw(T_r(H)) \geq mw(T_{r-1}(H)) \geq (x+1)p=((r-\lceil log p \rceil-1)/2+1)p=
(r-\lceil log p \rceil+1)p/2$.

Assume that $r=\lceil log p \rceil$ and let us show the lower bound of $p$ on the matching width.
$T_r$ contains $2^{\lceil log p \rceil+1}-1 \geq 2^{log p+1}-1=2p-1 \geq p$
vertices. By construction, $H$ contains at least $2p$ vertices.
Consequently, for each ordering of vertices of $T_r$ we can
specify a prefix and a suffix of size at least $p^2$ (just choose a prefix
of size $p^2$). Let $V_1$ be the set of vertices that got to the prefix
and let $V_2$ be the set of vertices that got to the suffix. By 
Lemma \ref{mincase} there is a matching of size at least $p$ consisting of
edges between $V_1$ and $V_2$ confirming the lemma for the considered case.

Let us now prove the lemma for $r=\lceil log p \rceil+2x$
for $x \geq 1$. Specify the centre of $T_r$ as the root  
and let $T^1, \dots, T^4$ be the subtrees of $T_r$ rooted by the grandchildren of the
root. Clearly, all of $T^1, \dots, T^4$ are copies of $T_{r-2}$.
Let $SV$ be a sequence of vertices of $V(T_r(H))$.  Let $SV^1, \dots, SV^4$
be the respective sequences of $V(T^1(H)), \dots, V(T^4(H))$ `induced' by $SV$
(that is their order is as in $SV$).
By the induction assumption, for each of them we can specify a partition $SV^i_1,SV^i_2$ into a prefix and
a suffix of size at least $p^2$ each witnessing the
conditions of the lemma for $r-2$. Let $u_1, \dots, u_4$ be the last respective vertices
of $SV^1_1, \dots, SV^4_1$. Assume w.l.o.g. that these vertices occur in $SV$ in the
order they are listed. Let $SV',SV''$ be a partition of $SV$ into a prefix and 
a suffix such that the last vertex of $SV'$ is $u_2$. By the induction assumption we know
that the edges between $SV^2_1 \subseteq SV'$ and $SV^2_2 \subseteq SV''$ form a matching $M$ of size at least $xp$.
In the rest of the proof, we are going to show that the edges between $SV'$
and $SV''$ whose ends do not belong to any of $SV^2_1,SV^2_2$ can be used to form a matching $M'$
of size $p$. The edges of $M$ and $M'$ do not have joint ends, hence this will imply
existence of a matching of size $xp+p=(x+1)p$, as required. 

The sets $SV' \setminus SV^2_1$ and $SV'' \setminus SV^2_2$ partition
$V(T_r(H)) \setminus (SV^2_1 \cup SV^2_2)=V(T_r(H)) \setminus V(T^2(H))=
V([T_r \setminus T^2](H))$. Clearly, $T_r \setminus T_2$ is a tree.
Furthermore, it contains at least $p$ vertices. Indeed, $T^2$ (isomorphic to $T_{r-2}$) has at least $p$
vertices just because we are at the induction step and $T_r$ contains at least
$4$ times more vertices than $T^2$. 
So, in fact, $T_r \setminus T^2$ contains at least $3p$ vertices. 
Furthermore, since $u_1$ precedes $u_2$, the whole $SV^1_1$ is in $SV'$.
By definition, $SV^1_1$ is disjoint with $SV^2_1$ and hence it is a subset
of $SV' \setminus SV^2_1$. Furthermore, by definition,
$|SV^1_1| \geq p^2$ and hence $|SV' \setminus SV^2_1| \geq p^2$ as well.
Symmetrically, since $u_3 \in SV''$, we conclude that $SV^3_2 \subseteq SV'' \setminus SV^2_2$
and due to this $|SV'' \setminus SV^2_2| \geq p^2$.

Thus $SV' \setminus SV^2_1$ and $SV'' \setminus SV^2_2$ partition  
$V([T_r \setminus T^2](H))$ into classes of size at least $p^2$ each and the size
of $T_r \setminus T^2$ is at least $3p$. Thus, according to Lemma \ref{mincase},
there is a matching $M'$ of size at least $p$ created by edges between 
$SV' \setminus SV^2_1$ and $SV'' \setminus SV^2_2$, confirming the lemma,
as specified above. $\blacksquare$

{\bf Proof of Theorem \ref{dmwtw}.}
First of all, let us identify the class ${\bf G}$. Recall that $P_x$ a path of $x$ vertices. 
Further on, let $0 \leq y \leq 3$ be such that
$k-y+1$ is divided by $4$.
The considered class ${\bf G}$ consists of all $G=T_r(P_{\frac{k-y+1}{2}})$ for $r \geq 5\lceil log k \rceil$.

Let us show that the treewidth of the graphs of ${\bf G}$ is bounded by $k$. 
Consider the following tree decomposition of $G=T_r(H=P_{\frac{k-y+1}{2}})$. The decomposition tree is $T_r$.
Consider $T_r$ as the rooted tree with the centre being the root. The bag of each vertex includes
the vertices of the copy of $H$ associated with this vertex plus the copy of the parent (for a non-root vertex). 
The properties of tree decomposition can be verified by a direct inspection.
The size of each bag is at most $k-y+1$, hence the treewidth is at most $k-y \leq k$.

Observe that max-degree of the graphs of ${\bf G}$ is $5$. Indeed, consider a vertex
$v$ of $G \in {\bf G}$ that belongs to a copy of $H$ associated with a vertex $x$ of some $T_r$.
Inside its copy of $H$, $v$ is adjacent to at most $2$ vertices. Outside its copy
of $H$, $v$ is adjacent to vertices in the copies of $H$ associated with the neighbours
of $x$, precisely one neighbour per copy. Vertex $x$ is adjacent to at most $3$ vertices
of $T_r$. It follows that $v$ has at most $3$ neighbours outside its copy of $H$.

In the rest of the proof we assume that $k \geq 50$. Te assumption does not restrict
generality because the constant can be made larger to incorporate smaller values of $k$.
Let us reformulate the lower bound of $mw(G)$ in terms of $log n$ and $k$ where $n=V(G)$.
Notice that $p$ used in Lemma \ref{dmwtwstruct} can be expressed as $(k-y+1)/4$.
Hence, the lower bound on the matching width can be seen as
$(r-\lceil log(\frac{k-y+1}{4}) \rceil+1)*(k-y+1)/8$. This lower bound can be immediately simplified
by noticing that by the choice of $k$ and $y$, $(k-y+1)/8 \geq k/16$ 
and $\lceil log(\frac{k-y+1}{4}) \rceil \leq \lceil log k \rceil$. Hence,
$(r-\lceil log k \rceil+1)k/16$ can serve as a lower bound on $mw(G)$. 
To draw the connection between $n$ and $r$, notice that $n=(2^{r+1}-1)(k-y+1)/2$.
It follows that $r+1=log(\frac{n}{(k-y+1)/2}+1)$. In particular, it follows that
$r+1 \geq log n-log k \geq log n -\lceil log k \rceil$. 
It follows that $r+1$ in the lower bound can be replaced by $log n -\lceil log k \rceil$ and the new
lower bound is $(log n-2 \lceil log k \rceil)k/16$. Consequently, for $log n \geq 5 \lceil log k \rceil$
the lower bound can be represented as $(log n*k)/32$ which is the form needed for the theorem.
It remains to observe that $r \geq 5\lceil log k \rceil$ implies $log n \geq 5\lceil log k \rceil$.
By the above reasoning, $r \geq 5\lceil log k \rceil$ implies $log(\frac{n}{(k-y+1)/2}+1) \geq 5\lceil log k \rceil$.
By our choice of $k \geq 50$, $log(n/20+1) \geq log(\frac{n}{(k-y+1)/2}+1)\geq 5\lceil log k \rceil$. By construction of $G$ and the choice of $r$,
$n \geq 2^{r+1}-1 \geq k^5-1 \geq k$, the last inequality follows from the choice of $k$, hence $n \geq 50$.
In particular, it follows that $n \geq n/20+1$. Hence $log n \geq log(n/20+1)\geq 5\lceil log k \rceil$.
$\blacksquare$

\section{Separation between {\sc robp} and decision-{\sc dnnf}}
\begin{lemma} \label{separ}
The space complexity of {\sc nrobp} on {\sc cnf} formulas $\phi(T_r(P_r))$
is $\Omega(n^{\log n/c})$ for some universal constant $c$.
\end{lemma}

{\bf Proof.}
The number of variables of $T_r(P_{2r})$ is $n=(2^{r+1}-1)*2r=2^r*4r-2r$.
That is, $r=\log \frac{n+2r}{4r} \geq \log n- \log r-2$. 
For a sufficiently large $r$, $r \geq \log r +2$, hence 
$r \geq \log n -r$ and hence $ r \geq \log n/2$.

By Lemma \ref{dmwtwstruct},
$mw(T_r(P_{2r})) \geq (r+1-\lceil log r \rceil)r/2$.
That is, $mw(T_r(P_{2r})) \geq (\frac{\log n}{2}+1-\log \frac {\log n}{2} -1)*\frac{log n}{4}$
It is not hard to see that for a sufficiently large $r$ (and hence sufficiently large $n$),
$mw(T_r(P_{2r})) \geq \frac{\log^2 n}{16}$. 
The statement of the theorem now follows immediately from Theorem \ref{nrobplbdmw}.
$\blacksquare$

\begin{theorem} \label{separ2}
There is an infinite class of {\sc cnf} formulas such that the complexity of decision-{\sc dnnf} on this
class is $O(n^5)$ while the complexity {\sc robp} is $\Omega(n^{\log n/c})$ for some universal 
constant $c$.
\end{theorem}

{\bf Proof.}
Consider the class $\phi(T_r(P_r))$. As a {\sc robp} can be seen as a special case of
an {\sc nrobp}, the lower bound on the space complexity of {\sc robp} on $\phi(T_r(P_r))$
immediately follows from Lemma \ref{separ}. 

It follows from Theorem 1 in \cite{DesDNNF} that the space complexity of decision-{\sc dnnf}
on a {\sc cnf} formula with primal graph treewidth $t$ is $O(2^tn)$ (the theorem in fact uses a different
parameter of a {\sc cnf} formula, however it is shown to never exceed the primal graph treewidth).
Arguing as in the proof of Theorem \ref{dmwtw}, we observe that the treewidth of $T_r(P_{2r})$
is at most $4r$.
We know from the proof of Lemma \ref{separ} that $r=\log \frac{n+2r}{4r}$. That is $r \leq \log(n+2r)$
and, for a sufficiently large $r$, $r \leq \log(2n)=\log n+1$. That is, for a sufficiently large
$r$, the treewidth of $T_r(P_{2r})$ is at most $4\log n+4$. Substituting $4\log n+4$ instead $t$ in $O(2^tn)$
results in $O(n^5)$, completing the required separation. $\blacksquare$

Thus, Theorem \ref{separ2} shows that the quasi-polynomial upper bound on the size of {\sc robp} 
simulating the given decision-{\sc dnnf} as described in \cite{BeameDNNF} is essentially tight.

\appendix
\section{Transformation of an {\sc nrobp} into a uniform one} \label{uniform}
Let $Z$ be a {\sc nrobp}. The \emph{in-degree} $d^{+}(v)$ of a node $v$ is a number of in-neighbours (this is essential
point because of the possibility of multiple edges). We assume that all the in-coming edges of nodes $v$ with
$d^{+}(v)>1$ are unlabelled. We call such a {\sc nrobp} \emph{clean}.
This assumption does not restrict generality because a {\sc nrobp} can be 
transformed into a clean one having at most twice more edges than the original {\sc nrobp}.
Indeed, let $v$ be a node with in-degree greater than $1$ and let $(u,v)$ be an edge labelled with 
a literal $x$. Subdivide $(u,v)$ and let $u,w,v$ be the path that replaced $(u,v)$. Then label $(u,w)$ with $x$.
Clearly, as a result we get a {\sc nrobp} implementing the same function as the original one. 
Notice that $w$ has in-degree $1$, that is the number of edges violating the assumption has decreased by $1$.
Thus, one can inductively argue that in case of $q$ `violating' edges, there is a transformation to a {\sc nrobp}
satisfying the above assumption that creates at most $q$ additional edges. 

For a node $v$ of $Z$, denote by $IVar_Z(v)$ the set of variables $x$ such that a literal of $x$ occurs on path
from the root to $v$ (the subscript can be omitted if clear from 
the context). We call the edges $(u,v)$ of $Z$ such that $d^{+}(v)>1$ \emph{relevant}.
A relevant edge $(u,v)$ \emph{irregular} if 
$IVar(u) \subset IVar(v)$ and \emph{regular} otherwise. 
Let $IVar(v) \setminus IVar(u)=\{x_1, \dots, x_q\}$.
Transform $Z$ as follows.
\begin{enumerate}
\item Remove the edge $(u,v)$.
\item Introduce new vertices $u_1, \dots, u_q$; we will refer to $u$ as $u_0$ for
the sake of convenience.
\item For each $1 \leq i \leq q$, introduce two edges $(u_{i-1},u_i)$ and label
them them $x_i$ and $\neg x_i$, respectively.
\item Introduce an unlabelled edge $(u_q,v)$.
\end{enumerate}

Let $Z'$ be the graph obtained as a result of the above transformation.
\begin{observation} \label{easyfacts}
\begin{enumerate}
\item $IVar_Z(v)=IVar_{Z'}(v)$.
\item The edge $(u_q,v)$ is regular in $Z'$.
\item $Z'$ is clean. 
\end{enumerate}
\end{observation}

{\bf Proof.} Immediate by construction. $\blacksquare$

\begin{lemma} \label{remainsnrobp}
$Z'$ is a {\sc nrobp} that computes the same function as $Z$.
\end{lemma}

{\bf Proof.}
To establish the read-once property of $Z'$,
it is sufficient to prove that any root-leaf path $P$ of $Z'$
that is not a path of $Z$ is read-once.
By construction, such a path $P$ includes $u$ and $v$
and the subpath $P_{u,v}$ starting at $u$ and ending at $v$ goes
through $u_1, \dots, u_q$ as defined above.
Let $P_u$ be the prefix of $P$ ending at $u$,
and $P_v$ be the suffix of $P$ beginning at $v$.
Notice that $P_u \cup P_v$ is a subgraph of a path of $Z$ and hence
cannot have repetitions of variable occurrences. 
By construction, $P_{u,v}$ does not have repeated variable occurrences either. 
A variable of $P_{u,v}$  does
not occur on $P_u$ because, by construction, the variables occurring
on $P_{u,v}$ do not belong to $IVar_Z(u)$. Finally all the variables occurring
on $P_{u,v}$, by construction, belong to $Ivar_Z(v)$ and hence cannot belong
to $P_v$. Indeed, otherwise if such a variable $x$ is found then there is a
path $P'$ of $Z$ from the root to $v$ on which $x$ occurs and hence $x$
occurs twice on $P'+P_v$ in contradiction to the read-once property of $Z$.
Thus we conclude that $Z'$ is indeed read-once. 

Let $S$ be a satisfying assignment of the function computed by $Z$
and let $P$ be a root-leaf path of $Z$ with $A(P) \subseteq S$.
If $P$ does not include $(u,v)$ then $P$ is a root-leaf path of $Z'$.
Otherwise, let $P_u$ and $P_v$ be as in the previous paragraph 
and let $P'$ be a $u-v$ path with $u_1, \dots, u_q$ being the intermediate vertices
and the in-edge for each $u_i$ is the one labelled with the literal of $x_i$ that
belongs to $S$ (by construction, such a selection is possible) and, as a result $A(P') \subseteq S$.
Taking into account that $A(P_u) \cup A(P_v) \subseteq A(P) \subseteq S$,
we conclude that $A(P_u+P'+P_v) \subseteq S$.
That is, in any case there is a root-leaf path of $Z$ whose set of literals is a subset of $S$ and hence 
$S$ is a satisfying assignment of the function computed by $Z'$.

Conversely, let $S$ be a satisfying assignment of the function computed by $Z'$.
Let $P$ be a root-leaf path of $Z'$ such that $A(P) \subseteq S$. If $P$ is not
a path of $Z$ then, by construction, $P$ includes both $u$ and $v$ and a path of
$Z$ can be obtained by replacement of the subpath of $P$ between $u$ and $v$ by an
edge $(u,v)$. Clearly, the set of literals of this resulting path is a subset of $A(P)$,
hence $S$ is a satisfying assignment of the function computed by $Z$.
$\blacksquare$

\begin{lemma} \label{moreregular}
The number of irregular edges of $Z'$ is smaller than the number of irregular edges of
$Z'$.
\end{lemma}

{\bf Proof.}
Denote by $Rel_{Z},Rg_Z,Rel_{Z'},Rg_{Z'}$ the sets of relevant edges of $Z$, regular edges of $Z$,
relevant edges of $Z'$, and regular edges of $Z'$, respectively. 
It is not hard to see that by construction, $Rel_{Z'}=(Rel_Z \setminus \{(u,v)\}) \cup \{(u_q,v)\}$.
That is, $|Rel_{Z'}|=|Rel_Z|$. By assumption, $(u,v) \notin Rg_Z$. Hence
$Rg_Z \subseteq Rel_{Z'}$. In fact $Rg_Z \subseteq Rg_{Z'}$. To show this, we need the following claim.

\begin{claim}
For each node $w \in V(Z) \cup V(Z')$, $IVar_Z(w)=IVar_{Z'}(w)$.
\end{claim}

{\bf Proof.}
Let $x \in IVar_Z(w)$ and let $P$ be a path from the root of $Z$ to $w$
containing an occurrence of $x$.
Note that by construction $P$ is either a path of $Z'$ or it can be replaced
by a path $P'$ with $A(P) \subseteq A(P')$. Hence $x \in IVar_{Z'}(w)$.

Conversely, let $x \in IVar_{Z'}(w)$ and let $P$ be a path from the root of $Z'$
to $w$ containing an occurrence of $x$. If $P$ does not contain $v$ then $P$ is path of $Z$
and hence $x \in IVar_Z(w)$. If $P$ contains $v$ but $x$ occurs on the suffix $P_v$ of $P$
starting at $v$ then, since $P_v$ is a path in $Z$, appending $P_v$ to an arbitrary
path from the root to $v$ will give us a path of $Z$ on which $x$ occurs. 
Finally if $x$ occurs on the prefix of $P$ ending at $v$ then $x \in IVar_{Z'}(v)$.
By the first statement of Observation \ref{easyfacts}, $IVar_Z(v)=IVar_{Z'}(v)$. That is, there is a path $P'$ of $Z$
from the root to $w$ that contains an occurrence of $x$. Consequently $P'+P_v$ is a a path
of $Z$ containing an occurrence of $x$. 
$\square$

Now, let $(u',v') \in Rg_Z$, that is $IVar_Z(u)=IVar_Z(v)$.
By the above claim, $IVar_{Z'}(u)=IVar_Z(u)=IVar_Z(v)=IVar_{Z'}(v)$.
That is, $(u',v') \in Rg_{Z'}$. 
Thus $Rg_{Z'}$ includes all the elements of $Rg_Z$ and, in addition
$(v_q,u)$, by the second statement of Observation \ref{easyfacts}. 
It follows that $|Rg_{Z'}|>|Rg_Z|$.
Now, the number of irregular edges of $Z$ and $Z'$ are, respectively,
$|Rel_Z|-|Rg_Z|$ and $|Rel_{Z'}|-|Rg_{Z'}|$. It follows from the proved
above that the latter is smaller than the former. $\blacksquare$

\begin{theorem}
Let $Z$ be a clean {\sc nrobp} with $q$ irregular edges.
Then there is a uniform {\sc nrobp} $Z^*$
computing the same function as $Z$ and having at most $2qn$ edges 
more than $Z$.
\end{theorem}

{\bf Proof.}
By induction on $q$. If $q=0$ then all the relevant edges of $Z$
are regular. It is easy to observe that in this case $Z$ is uniform
and hence no further transformation is needed.

Suppose $q>1$. Pick an irregular edge $(u,v)$ and transform $Z$ to $Z'$
as specified above. By Lemma \ref{remainsnrobp}, $Z'$ is a {\sc nrobp}. 
By the third statement of Observation \ref{easyfacts}, $Z'$ is clean.
By Lemma \ref{moreregular}, $Z'$ has at most $q-1$ irregular edges.
Hence, by the induction assumption, there is a uniform {\sc nrobp} $Z^*$
computing the same function as $Z'$ and having at most $2(q-1)n$ more edges
than $Z'$. As $Z'$ computes the same function as $Z$, by Lemma \ref{remainsnrobp}
and, by construction, has at most $2n$ edges more than $Z^*$, we conclude that
$Z$ computes the same function as $Z$ and has at most $2qn$ more edges.
$\blacksquare$
 
\section{Equivalence of the {\sc arosrn} and the traditional definition of the {\sc nrobp}}
A ({\sc nrobp}) is traditionally defined as a {\sc dag} $Z$
with one root and two leaves. Some of non-leaf nodes are labelled with variables so that 
no variable occurs as a label twice on a directed path of $Z$. A node labelled with a variable has two outgoing
edges one labelled with $true$ the other with $false$. Finally, the leaves are labelled with
$true$ and $false$. 

It is convenient to see each edge $e$ labelled with $true$ or $false$ being in fact
respectively labelled with the positive or negative literal of the variable 
labelling the tail of $e$. With such a labelling an assignment $A(P)$ associated with
each directed path of $Z$ is simply the set of literals labelling the edges of $P$.
The satisfying assignments of the function computed by $Z$ are precisely those that are extensions
of $A(P)$ for paths $P$ from the root to the $true$ leaf.

It is not hard to see that for any function that is not constant $false$,
{\sc nrobp} can be thought as a special case of {\sc arosrn}. Indeed, with edges
labelled by literals as specified in the previous paragraph, remove the labels
from the vertices, remove the $false$ leaf as well as all nodes of $Z$ from which th
$true$ leaf is not reached and the obtained graph is an {\sc arosrn} computing exactly
the same function as $Z$.

Conversely, an {\sc arosrn} can be transformed into a {\sc nrobp} as follows.
Denote the only leaf of the {\sc arosrn} as the $true$ leaf and introduce a new node to be
the $false$ leaf. Then for each edge $(u,v)$ labelled with a literal $x$, apply the following
transformation.
\begin{itemize}
\item Subdivide $(u,v)$ by introducing a new node $w$ and edges $(u,w)$ and $(w,v)$
instead $(u,v)$.
\item Introduce a new edge $e$ from $w$ to the $false$ leaf.
\item Label $w$ by $Var(x)$, the variable of $x$.
\item If $x$ is the positive literal then label $(w,v)$ with $true$ and $e$ with $false$.
Otherwise, label $(w,v)$ with $false$ and $e$ with $true$.  
\end{itemize}

\begin{figure}[h]
\includegraphics[height=4cm]{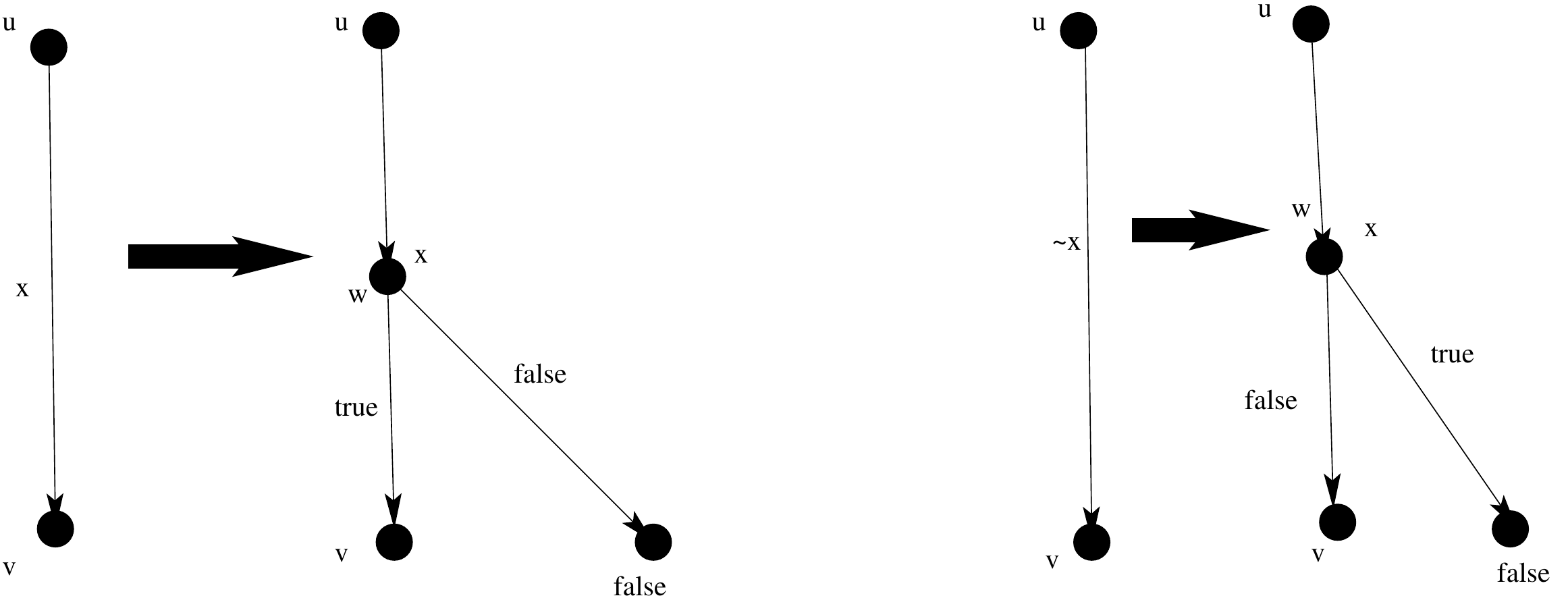}
\caption{Transformation of a labelled edge of an {\sc arosrn}.}
\label{arosrntransf}
\end{figure}

The transformation of labeled edges is illustrated in Figure \ref{arosrntransf}.

It is not hard to see that there is a bijection between root-leaf paths of the {\sc arosrn}
and root-true leaf paths of the resulting {\sc nrobp} preserving the associated sets of literals.
Therefore, we conclude that this transformation is valid.
\end{document}